\documentclass[twocolumn]{openjournal}

\usepackage{xcolor}
\usepackage{textgreek}
\usepackage{amsmath}
\usepackage[utf8]{inputenc}
\usepackage[english]{babel}

\usepackage{hyperref}
\hypersetup{
    unicode, 
    colorlinks=true,
    linkcolor=linkcolor,
    citecolor=linkcolor,
    filecolor=linkcolor,
    urlcolor=linkcolor,
}
\usepackage{color,colortbl}
\definecolor{linkcolor}{rgb}{0.0,0.3,0.5}
\usepackage{tensind}
\tensordelimiter{?}
\DeclareGraphicsExtensions{.bmp,.png,.jpg,.pdf}
\usepackage{verbatim}
\usepackage[normalem]{ulem}

\def\be{\begin{equation}}
\def\ee{\end{equation}}

\def\d{{\rm d}}

\def\lsim{\raise0.3ex\hbox{$\;<$\kern-0.75em\raise-1.1ex\hbox{$\sim\;$}}}
\def\gsim{\raise0.3ex\hbox{$\;>$\kern-0.75em\raise-1.1ex\hbox{$\sim\;$}}}
\def\eps{\varepsilon}
\def\theta{\vartheta}

\renewcommand{\vec}[1]{\boldsymbol{#1}}

\usepackage{xcolor,color}
\definecolor{darkBlue}{rgb}{0, 0, 0.8}

\begin{document}

\title{Cygnus~X-3 as a PeVatron and the LHAASO 2025 data}

\author{M.~Kachelrie\ss$^{1}$}
\author{E.~Lammert$^{1,2}$}

\affiliation{$^{1}$Institutt for fysikk, NTNU, Trondheim, Norway}
\affiliation{$^{2}$Department of Physics, School of Natural Sciences,
  Technical University Munich, Germany}

\begin{abstract}
We have recently argued that the high-mass X-ray binary Cygnus~X-3
can accelerate cosmic rays (CR) beyond PeV energies. Meanwhile, the LHAASO
collaboration published the measurement of an orbitally modulated photon
flux from Cygnus~X-3 extending up to 4\,PeV. In this short extension of our
previous work, we argue that these observations
point towards CR acceleration in the jet, and secondary production in
CRs scattering on gas from the wind and on stellar UV photons from the
companion star. 
The latter channel leads naturally to a contribution to the photon flux
peaking around PeV energies which is strongly orbitally modulated.
The fast drop
in the flux of these photons below PeV energies may be caused by 
absorption on an increased density of background photons in a line-driven
stellar wind.
\end{abstract}


\maketitle

{\em Introduction\/}
The high-mass X-ray binary Cygnus~X-3 has been suggested since the 1980s as
a source of photons and neutrinos with energies up to the PeV 
range~\citep{1994ApJS...90..883P}. In a previous
work, \citet{Kachelriess:2025hkj} (KL25 hereafter) confirmed
that in this environment several CR acceleration mechanisms like  diffusive
shock acceleration,
2nd order Fermi acceleration and magnetic reconnection are able to
accelerate CRs to energies of tens of PeV. At that time, only 
measurements of the photon flux from the extended Cygnus region were
available in this energy range, and thus no  comparison of the predictions
of KL25 to data could be performed.

Meanwhile, the LHAASO collaboration published the measurement of a photon
flux extending up to 4\,PeV from Cygnus~X-3 which shows evidence for an
orbital modulation~\citep{LHAASO}. In this short addendum, we discuss how
these observations fit to the models employed by KL25.
First, we note that the spectral shape of
the measured photon flux shown in Fig.~1 points towards two components: one
with a flat\footnote{We refer always to the slope in an $E^2\d N/\d E$ plot.}
spectrum  as expected for hadronic interactions, and another one
rising at PeV energies.
This rise can be simplest interpreted as a threshold effect in
photo-hadronic interactions, and its peak energy is compatible with
interactions of CR protons on UV photons from the Wolf-Rayet (WR)
companion star.
The non-observation of a photon flux at the lowest energy
bin may be connected either with the onset of $\gamma\gamma$ internal
absorption, but is also consistent with a roughly flat photon spectrum.
Note also that the limits from the
MAGIC collaboration at lower energies are
less restrictive~\citep{2010ApJ...721..843A,TeVPA}.

The second piece of information is the evidence for an orbital modulation of
the photon flux, which is approximately in phase with the one seen at lower
energies in Fermi/LAT data~\citep{Abdo:2009kfa,Zdziarski:2018dms}.
This agreement suggests a
similar origin of the flux modulation in the GeV and PeV range, namely
the anisotropy of the UV photon field far away from the emitting
surface of the WR star. In the GeV
range, inverse Compton scattering of electrons on the stellar UV photons
in the jet was used to fit successfully the observed Fermi/LAT gamma-ray
orbital modulation of Cygnus X-3~\citep{Dub10}.
In the case of photo-hadronic interactions of ultra-relativistic protons,
the secondary mesons and their decay products are  emitted mainly
into the forward direction. Thus up to corrections of order
$\delta\theta\sim 1/(2\gamma)$, with $\gamma\sim 10^4-10^5$ as the Lorentz
factor of the initial proton, the 3-momenta of the proton, mesons
and photons are
aligned. Since moreover the jet direction $\vec e_{\rm jet}$ of Cygnus~X-3 is
constant during the orbital motion, also the photon and thus the parent
proton direction $\vec p$ is fixed, $\vec p\| \vec e_{\rm jet}$.
As a result, the relative angle  between the
3-momenta of the primary and the background photon will change during one
orbit, leading to a strong orbital modulation of the photon flux.

{\em Reaction rates\/}
We discuss next how we implement this anisotropy into the calculation
of the reaction rates.
For a strongly anisotropic background of photons with energy $\eps$ and
differential number density $n_{\gamma}(\eps)$, we approximate the reaction
rate $R$ of the primary $i$ as
\begin{align}  \label{rate1}
 R_i(E) & =   (1-\mu_0)\;\int_0^\infty \!\!\! \d\eps \; n_{\gamma}(\eps) 
 \sigma_{i\gamma}(s)\;\Theta(s-s_{\min}) ,
\end{align}
with a fixed relative direction $\mu_0\equiv\cos\theta_0$ between the
3-momenta of the primary nuclei and background photon, $\sigma_{i\gamma}$
the cross section, $\Theta(s)$ the Heaviside step function, and $s_{\min}$
the threshold energy squared. Such an approximation is justified at
sufficiently large distances compared to the extension of photon
source. In the case of reactions on stellar photons, 
both the squared  center-of-mass energy of the two particles,
$s=2E\eps(1-\mu_0)$, and the flux factor $1-\mu_0$ change with the angle
$\theta_0$ during one orbit,
leading to an orbital modulation of the reaction rate.
In contrast, the angle $\theta_0$ is fixed for reactions on photons from
the accretion disk, with angles restricted to $(1-\mu_0)\simeq (R/d)^2$.  
Thus for such reactions, no orbital modulation is expected but the threshold
energy is increased by the factor $(R/d)^2$ compared to an isotropic
background.

\begin{figure}
  \includegraphics[width=.49\textwidth]{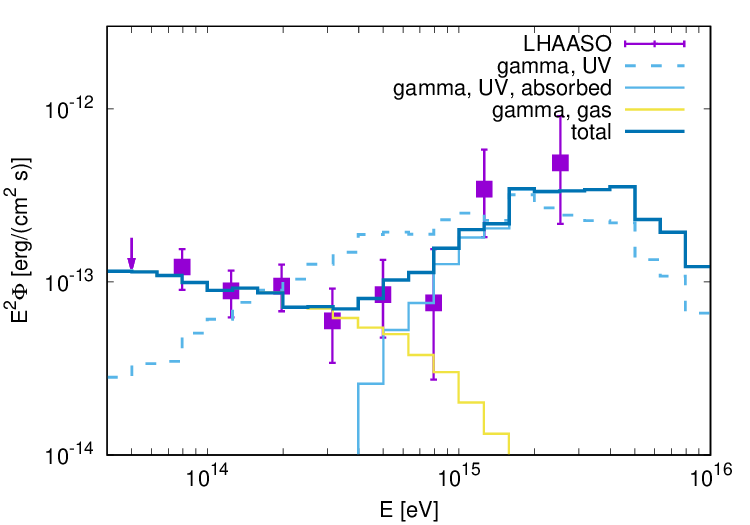}
  \caption{The average photon flux as function of energy compared to
    the LHAASO data~\citep{LHAASO} from Cygnus~X-3  in the flaring state
    (corrected for Galactic absorption); shown is also separately
    the contribution from scatterings on UV photons and gas.
  }
  \label{fig:spec}
\end{figure}

In addition to UV and accretion disk photons, X-ray photons are a potential
target for photo-hadronic interactions. KL25 assumed
that the observed X-ray flux is emitted as synchrotron radiation by electrons
accelerated in the same region as nuclei. Then the X-ray photons would be
isotropically distributed in the acceleration zone, and no orbital
modulation is expected in the reaction rate on
these background photons. Thus a dominant contribution of X-ray photons to the
interaction depth of CRs is disfavoured, if the evidence for an orbital
modulation seen in the LHAASO data is confirmed. Moreover, the modulation of
the X-ray data has a phase opposite to the data in the Fermi/LAT and
LHAASO energy
range, suggesting a different emission region of X-rays and high-energy
photons.
An emission  close to the black hole is also favoured by
X-ray observations in other micro-quasars, as well as by data from the
Imaging X-ray Polarimetry Explorer (IXPE) observational campaign
on Cygnus~X-3, which  attributes the X-ray emission to an optically thick
envelope close the central source~\citep{Veledina:2024xpr,Mikusincova:2025imj}.
We assume therefore, in contrast to our previous
work, that the reactions on these X-ray photons are negligible.

The only other change in the choice of parameters compared to KL25
concerns the strength of the magnetic field in
the acceleration
zone. If the main part of X-ray emission is decoupled from the acceleration
zone of hadrons, no strong observational constraints on the magnetic field
exist. To be conservative, we choose therefore a field strength
of the magnetic field, $B=2000$\,G, as low as being compatible with an
Hillas energy of $E=6\times 10^{16}$\,eV. Stronger magnetic fields
would lead also to $\gamma\to e^+e^-$ pair cascades, and are therefore
disfavoured by the observation of PeV photons.  As a result, acceleration
mechanisms relaying on strong magnetisation, like magnetic reconnection
or Fermi 2nd order acceleration, are less effective. Therefore, we consider
only diffusive shock acceleration as the acceleration mechanism in the jet.

\begin{figure}
  \includegraphics[width=.49\textwidth]{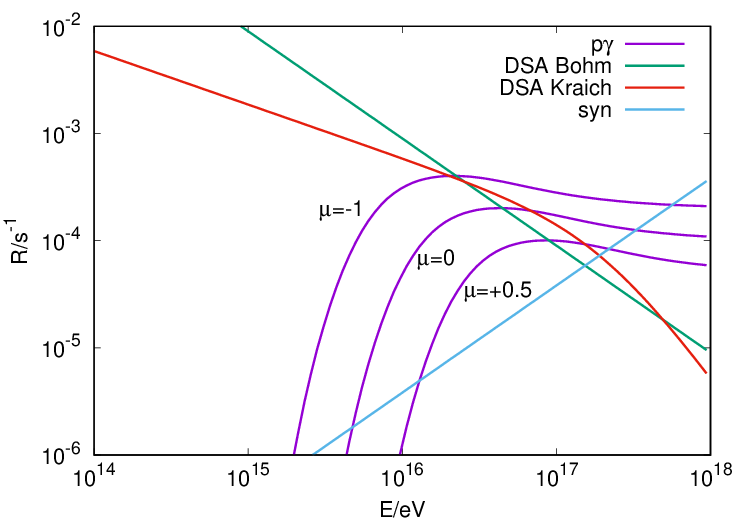}
  \caption{The energy gain rates for shock acceleration assuming Bohm and
    Kraichnan diffusion, as well as the energy loss rates due to synchrotron
    radiation and photo-hadronic interaction of stellar UV photons as function
    of the proton energy.
  }
  \label{fig:rates}
\end{figure}

{\em Numerical results}
In Fig.~\ref{fig:rates}, we show the resulting energy loss and gain rates
for protons.
For the  energy-loss rate $\tau^{-1}=1/E (\d E/\d t)$ due to photo-hadronic
interactions, we include in Eq.~(\ref{rate1}) the inelasticity $y$ of the
corresponding reaction. They are shown for three choices of the scattering
angle, $\mu=\cos\theta$, and exhibit the expected increase of the effective
threshold energy with decreasing values of $\mu$.
For the energy gain rate $R_{\rm DSA}= \xi v_{\rm sh}^2/D$ due to diffusive shock
acceleration~\citep{Lagage:1983zz}, we show the rate both for (isotropic)
Bohm and Kraichnan diffusion,
where the diffusion coefficient $D$ of a particle with Larmor radius
$R_{\rm L}$ is given by~\citep{Giacinti:2017dgt} 
\be  \label{diff}
  D = \frac{cL_0}{3} 
 \left[ (R_{\rm L}/L_0)^{2-\gamma} + (R_{\rm L}/L_0)^2 \right] 
\ee
with $\gamma=1$ for Bohm and $\gamma=3/2$ for Kraichnan turbulence,
respectively. Note that the transition energy between small- and large
angle scattering at $R_{\rm L}\simeq L_0$ corresponds to the Hillas energy,
if the typical length scale  $L_0$ of the turbulent field fluctuations is
comparable to the extension $D$ of the acceleration region, which we choose as
$L_0=D=7\times 10^{11}$\,cm. In this case, the maximal energy achievable
with Kraichnan diffusion is similar to the one in Bohm diffusion. For
simplicity, we use therefore in the following only Bohm diffusion.
In all cases, energies beyond 10\,PeV are reachable.

In Fig.~\ref{fig:spec}, we compare the measurements of LHAASO in the flaring
state to
the photon spectra obtained using the Monte Carlo framework
of KL25 averaged over one orbital period.
Except for the adjustments described above, the Monte Carlo framework
is unchanged and we refer for details to KL25.  Note
that we neglect the effect of photon  absorption during the propagation in
the Milky Way and compare to the  de-absorbed LHAASO data.
We show  the separate contributions produced in
interactions on the gas from the stellar wind (yellow, solid) and on UV photons (light blue, dashed): While the
former spectrum follows the CR slope, the latter is  rising $\propto 1/E$
until $\simeq $\,PeV energies. This rise, typical for photon fluxes below
the threshold energy if there is no strong beaming, is too slow to reproduce
well the measurements. As a possible explanation for the fast rise of
the measured photon flux, we
explore next the option that the photons produced in p$\gamma$ interactions
with energies below 1\,PeV are (partially) absorbed, i.e.\ that
$\tau_{\gamma\gamma}\simeq 1$ at 1\,PeV. This would require that the absorption
is increased by a factor of order 100-1000 compared to the standard
treatment of, e.g., \citet{2011A&A...529A.120C}: There, one
assumes for the density $n(\eps)$ of the background photons a Planck
distribution which is reduced by the geometrical factor $(R_s/d)^2$ which
is given by the radius $R_s$ and  the distance $d$ to the WR star.
In this picture,
stellar photons are free-streaming from the photosphere defined by
$\tau_{\rm es}=2/3$ using as opacity Thompson scattering on free electrons.
In the case of WR stars, the required efficient momentum transfer from
the photons to the wind is provided by the multiple absorption and
re-emission of photons by ionised gas in the wind. The region of the wind
where photon scattering on ions is efficient is limited by
the condition that the wind has not yet reached its terminal velocity
$v_\infty$. Approximating the sequence of multiple absorption and
re-emission of photons as a random walk, the number of scatterings
is $N\simeq (v_\infty/\Delta v)^2$, with $\Delta v\simeq 1$km/s as the
redshift spacing of the lines~\citep{Lamers99}. Relative to free-streaming, the
density of photons in a line-driven wind is thus enhanced by the factor
$(N)^{1/2}\simeq v_\infty/\Delta v\sim 2000$. \citet{Vilhu:2021vfs} found
that the wind of Cygnus~X-3 approaches its terminal velocity, depending on
the radiation state, within $\sim 3R_s$, a value we use as estimate for
the size of the region with enhanced absorption.
Note that the acceleration region is at larger distances, so that the
free-streaming approximation holds there.
In Fig.~\ref{fig:spec}, we show the photon flux including increased
absorption on 
background photons with solid lines. Clearly, the predicted
flux including enhanced absorption agrees better with the data.

Let us now check if the assumption $\tau_{\gamma\gamma}\sim 1$ at
$E=1$\,PeV is realistic.  We employ from the suite of the Potsdam Wolf-Rayet
models\footnote{The temperature profile can be downloaded
  from~\url{https://www.astro.physik.uni-potsdam.de/PoWR/}}
MW WNL-H20 13-19 as the numerical model for the line-driven wind of
Cygnus~X-3~\citep{2004A&A...427..697H,2015A&A...579A..75T}. In this model,
the temperature reaches 254.000\,K at the surface of the WR star, dropping
to 57.000\,K at $r=3R_S$. Moreover, the suppression factor
$(R_s/r)^2$ is absent as long as re-scattering is effective. As a result,
the number density of target photon is increased between a factor
$(254/40)^3\simeq 260$ and $3^2=9$.  Hence line-driven winds lead indeed to
increased photon absorption close the WR star. A confirmation of the strong
rise of the photon flux towards PeV energies would suggest that the
line-of-sight (LoS) of PeV photons has a small impact parameter to the WR
star, limiting thereby the location of the acceleration region of PeV
particles.

From a plot similar to Fig.~\ref{fig:rates} for electrons, one reads off
50\,GeV as the maximal
electron energy which is restricted by synchrotron losses. This corresponds
to a maximal synchrotron frequency in the MeV range, leading in turn to the
emission of photons up to 10\,TeV via inverse Compton scattering. Thus a
leptonic scenario can, for our parameter choices, not explain the PeV
flux observed by LHAASO.

\begin{figure}
  \includegraphics[width=.49\textwidth]{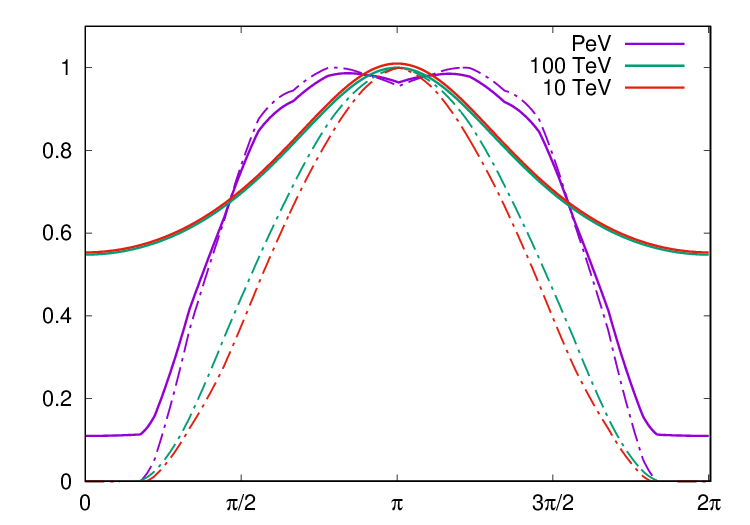}
  \caption{The modulation amplitude of the total photon flux
    (solid lines) and the photon flux produced in
    photo-hadronic interactions (dashed lines)
    as function of the orbital phase
    for three energies; note that the 10 and 100\,TeV curves for the
      total flux practically overlap.
      }
  \label{fig:mod}
\end{figure}

\begin{figure}
  \includegraphics[width=.49\textwidth]{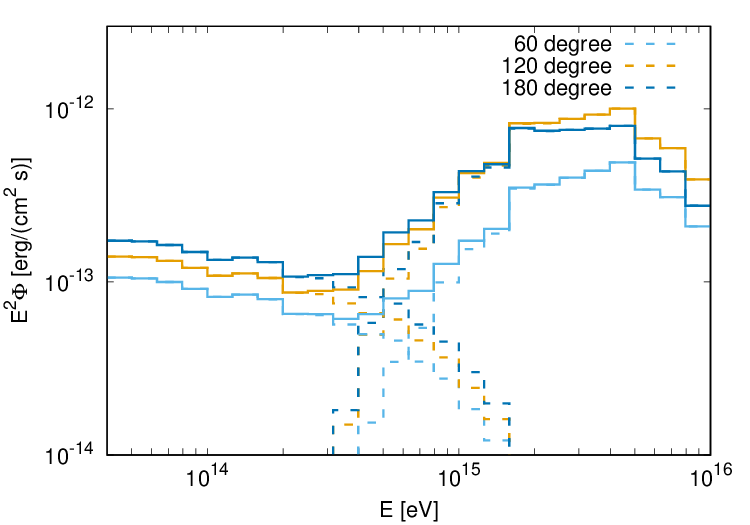}
  \caption{The photon flux as function of energy
      for three values of the orbital phase; dashed lines for individual
      contributions from interaction on gas and UV photons, solid lines for
      the total flux.
      }
  \label{fig:mod2}
\end{figure}

Finally, we show in Fig.~\ref{fig:mod} the expected modulation of the
photon flux produced by interactions on UV photons by dashed lines.
Here, we assume that
the bulk motion of the acceleration zone is moderately relativistic so that
the neglected contribution of the counter-jet is sufficiently suppressed --
an approximation justified by the large uncertainties of the modulation
measured by LHAASO. While at lower energies,
the modulation is close to the flux factor $(1-\cos\theta)/2$, at PeV energies
the shape is more complicated. Here, both the changes in the effective
threshold energy and the maximal energy visible in Fig.~\ref{fig:rates}
play a role. Next we consider the modulation of the photon flux produced in
interactions on gas, which results from the different LoS
integrals entering the interaction depth:
As a result, the photon flux
from interaction with gas is modulated by a factor two. The modulation
of the total photon flus is shown by solid lines in Fig.~\ref{fig:rates}.
  In addition, we show in Fig.~\ref{fig:mod2} the predictions for the
  modulated photon flux $E^2\Phi$ as function of the photon energy $E$
  for three orbital phases, $\phi=60^\circ,120^\circ$ and $180^\circ$.

{\em Conclusion}
The spectral shape of the photon flux measured by the LHAASO collaboration
is in rough  agreement with our prediction based on CR acceleration in the
jet and a two-component model based on interactions on gas and UV photons.
The sharp rise of the photon flux around PeV may indicate that the density
of UV photon close to the WR star is enhanced by multiple scattering on
ionised gas in the stellar wind. In such a model, one expects a strong orbital
modulation in the PeV energy range, while the modulation minimum is
decreasing at lower energies towards $\simeq 0.55$. If the evidence
for a modulation in future
data increases, including relativistic beaming effects may help to restrict
the jet geometry and parameters. Since such an orbital
modulation is also predicted for the neutrino flux, using  a
time template would increase the sensitivity of neutrino correlation studies.

{\em Acknowledgements}
We would like to thank Karri Koljonen for useful discussions and advice on the
Posdam WR models.


\bibliographystyle{aasjournal}

\bibliography{arxiv}

\end{document}